\documentclass[a4paper,12pt]{article}

\usepackage[height=8.85in,width=6.45in]{geometry}
\usepackage{fancybox}
\usepackage{amsmath,amssymb}
\usepackage[T1]{fontenc}
\usepackage{comment}
\usepackage{cite}
\usepackage{hyperref}
\usepackage{xcolor}

\def\a{\alpha}

\def\aa{{\dot \a}}

\def\CC{{\cal C}}

\def\CL{{\cal L}}

\def\CN{{\cal N}}

\usepackage{slashed}

\def\Pin{\mathop{\rm Pin}}
\def\pin{\mathop{\rm pin}}
\def\SO{\mathop{\rm SO}}

\def\SU{\mathop{\rm SU}}
\def\U{\mathrm{U}}

\def\SL{\mathop{\rm SL}}
\def\tr{\mathop{\rm tr}}

\def\CC{{\cal C}}

\def\CL{{\cal L}}

\def\CN{{\cal N}}

\newcommand{\bR}{\mathbb{R}}
\newcommand{\bZ}{\mathbb{Z}}

\def\SU{\mathrm{SU}}

\def\U{\mathrm{U}}
\def\SO{\mathrm{SO}}

\def\SL{\mathrm{SL}}
\def\tr{\mathop{\mathrm{tr}}\nolimits}

\def\cH{{\cal H}}

\def\sT{\mathsf{T}}

\def\sR{\mathsf{R}}

\def\time{\sT}
\def\reflection{\sR}

\def\beq#1\eeq{\begin{align}#1\end{align}}

\usepackage{framed}
\definecolor{shadecolor}{rgb}{0.90,0.90,0.90}

\usepackage{times}
\usepackage{courier}
\usepackage{mathtools}
\numberwithin{equation}{section}

\let\bar\overline

\def\CC{\mathbb{CC}}
\def\MO{\mathbb{MO}}
\def\dimreduce{{\textstyle\int\nolimits}}

\newcommand{\vev}[1]{ \left\langle {#1} \right\rangle }

\newcommand{\ket}[1]{ | {#1} \rangle }

\begin{document} 

\begin{titlepage}

\begin{flushright}
IPMU-16-0129
\end{flushright}

\vfill

\begin{center}

{\Large\bfseries On time-reversal anomaly of 2+1d topological phases}

\vskip 1cm
Yuji Tachikawa and Kazuya Yonekura
\vskip 1cm

\begin{tabular}{ll}
  & Kavli Institute for the Physics and Mathematics of the Universe, \\
& University of Tokyo,  Kashiwa, Chiba 277-8583, Japan
\end{tabular}

\vskip 1.5cm

\textbf{Abstract}

\end{center}

\vskip1cm

\noindent
We describe a method to find the anomaly of the time-reversal symmetry of  2+1d topological quantum field theories, by computing the fractional anomalous momentum on the cross-cap background.
This allows us, for example, to identify the parameter $\nu$ mod 16 of the bulk 3+1d topological superconductor with $\time^2=(-1)^F$ on whose boundary a given 2+1d time-reversal-invariant topological phase can appear. 

\vfill

\end{titlepage}
\tableofcontents

\section{Introduction and summary}
A quantum field theory in $d+1$ spacetime dimensions with a global symmetry $G$ can have an anomaly.
This anomaly manifests itself as the phase ambiguity of its partition function in the presence of a nontrivial background gauge field for the global symmetry $G$.
Moreover, this phase ambiguity appears in a controlled manner. 
For example, when $G$ is a continuous internal symmetry, it follows the Wess-Zumino consistency condition.

More generally, the phase ambiguity can be understood by regarding the original quantum field theory as realized \emph{on the boundary} of another quantum field theory in $(d+1)+1$ spacetime dimensions with the same global symmetry $G$ \emph{in the bulk}.
The bulk theory is almost trivial in the sense that there is a unique gapped vacuum on any compact spatial manifold, and is known under various names, such as an invertible field theory in the math literature or as a symmetry protected topological phase (SPT phase) in the condensed matter literature. 
When the spatial manifold has a boundary, the partition function of the total system is  properly $G$ symmetric, since the phase ambiguity of the boundary theory is canceled by the phase of the bulk system.
In this manner, the bulk SPT phase completely captures the anomaly of the boundary theory.
This mechanism is long known as the anomaly inflow when the symmetry $G$ is continuous.

In this paper, we study the anomaly of the time-reversal symmetry $\time$ of 2+1d relativistic quantum field theories with fermions, such that $\time^2=(-1)^F$.
The corresponding bulk theory is known as 3+1d topological superconductors with $\time^2=(-1)^F$, and has received much attention in the recent years.
There are now ample pieces of evidence that such topological superconductors follow a $\bZ_{16}$ classification \cite{Fidkowski:2013jua,Wang:2014lca,KitaevCollapse,Metlitski:2014xqa,Morimoto:2015lua,Seiberg:2016rsg,Tachikawa:2016xvs,Witten:2016cio}.
Correspondingly, given a time-reversal-symmetric 2+1d system with $\time^2=(-1)^F$,
its time-reversal anomaly is characterized by an element of $\bZ_{16}$~\cite{Witten:2015aba,Hsieh:2015xaa}.
For example, the anomaly of a system with $\nu$ massless 2+1d Majorana fermions is given by $\nu$ mod 16.
With interactions, the boundary theory can also be gapped and become a topological quantum field theory (TQFT).
The main objective of this paper is to provide a method to compute the $\bZ_{16}$ anomaly of the time-reversal symmetry when the 2+1d system is topological.

For this purpose, we first translate  the $\bZ_{16}$ anomaly to the fractional background momentum on the crosscap background.
Note first that introducing a background gauge field for the time-reversal symmetry amounts to considering the theory on a general non-orientable manifold.
Let us therefore put the 2+1d system, here not assumed to be topological, on the spatial geometry with the coordinates $(x,\theta)$, with the identification \begin{equation}
(x,\theta)\sim (-x,\theta+\pi).\label{ccgeometry}
\end{equation}
One may see this manifold as a half cylinder $x \geq 0,~\theta \sim \theta +2\pi$ with the identification condition $(0, \theta) \sim (0, \theta+\pi$)
which can be recognized as a crosscap.
This background has a $\U(1)$ isometry shifting $\theta$, such that $\theta\mapsto\theta+2\pi$ is an identity. 
On a system without anomaly, the corresponding momentum is integrally quantized.
We will see below that, on a system with anomaly, we instead have \begin{equation}
p=n + \frac{\nu}{16}, \qquad n\in \bZ,\label{anomalousmomentum}
\end{equation}i.e.~the background has an anomalous momentum. 
More precisely, there are two complementary pin$^+$ structures on the crosscap geometry, and the momentum on one is given by \eqref{anomalousmomentum} and the momentum on the other is given by $p=n-\nu/16$. 

To determine the anomaly of a 2+1d TQFT, then, we need to compute this anomalous momentum. We will see below that this can be done by determining the state on a torus $T^2$ created by the crosscap times a circle $S^1$. 

Before proceeding, we note that essentially the same method to understand the anomaly of the time-reversal symmetry of 1+1d systems with 2+1d bulk SPTs in terms of the anomalous fractional momentum has already been discussed in \cite{Cho:2015ega}. The novelty here is to apply it to 2+1d systems with 3+1d bulk SPTs, in particular to 2+1d topological systems.
We also note that in \cite{Chan:2015nea} the authors already considered a particular class of 2+1d topological theories, namely the Abelian Chern-Simons theories, on non-orientable manifolds, but they only considered non-anomalous theories.

The rest of the paper is organized as follows.
In Sec.~\ref{why}, we first explain why the relation \eqref{anomalousmomentum} holds, using some  general argument and an explicit example of $\nu$ Majorana fermions. 
In Sec.~\ref{how}, we describe how we can determine this anomalous momentum in the case of 2+1d topological systems. 
In Sec.~\ref{examples}, we apply the methods developed in Sec.~\ref{how} to a couple of examples.
We also have an appendix where we realize the semion-fermion theory on the domain wall of a super Yang-Mills theory.

We pause here to mention that to discuss 2+1d TQFTs on non-orientable manifolds properly, we should first generalize the standard Moore-Seiberg axiom for 2+1d orientable TQFTs \cite{Moore:1988qv,Moore:1989vd} to the non-orientable situation. 
The authors plan to do so in the future; in the present paper, we develop only the minimal amount of machinery necessary to determine the time-reversal anomaly.\footnote{
We also remark that in this paper we use comparison of TQFTs with fermions where the relation between anomalies and SPT phases (see e.g.,
\cite{Wen:2013oza,Wang:2013yta,Wang:2014pma,Kapustin:2014lwa,Kapustin:2014zva,Hsieh:2015xaa}) are
well-established for fermions~\cite{Witten:2015aba} by the Dai-Freed theorem~\cite{Dai:1994kq,Yonekura:2016wuc}. 
It would be very interesting to find the right mathematical structure directly in TQFTs without relying on 
the anomaly matching with fermions.}

\section{Time-reversal anomaly and the anomalous momentum}\label{why}
Consider the crosscap geometry \eqref{ccgeometry} as in the introduction.
Let $P_\theta$ be the momentum operator in the direction $\theta$. Then, $e^{2 \pi i P_\theta}$ is
a trivial element of the symmetry group in non-anomalous theories. However, in anomalous theories, this can be nontrivial.

The reason that $e^{2 \pi i P_\theta}$ corresponds to an anomaly is as follows. 
Suppose that we want to compute the thermal partition function $\tr e^{-\beta H}$ in the crosscap geometry (with the infinite spatial volume appropriately regularized). 
If we insert $e^{2 \pi i P_\theta}$ and compute
$\tr e^{- \beta H+2 \pi i P_\theta}$, it is just computing the path integral on the geometry $\text{crosscap} \times S^1$ which is completely the same
manifold as for the path integral expression for $\tr e^{-\beta H}$. However, the results change by the phase factor $e^{2\pi i p }$, where $p \in \bR /\bZ$
is the anomalous momentum. The phase ambiguity of partition functions is an anomaly. 
This is very similar to the explanation of the anomaly of the modular invariance under the element $T \in \SL(2,\bZ)$ in 1+1d field theories.

Suppose we have two theories with the values of the time reversal anomaly $\nu \in \bZ_{16}$ 
given by $\nu_1$ and $\nu_2$ and the anomalous momentum given by $p_1$ and $p_2$.
The time reversal anomaly and the anomalous momentum are additive quantities;
if we consider the theory which is a direct product of the two theories, then the time reversal anomaly and the anomalous momentum are simply given by
$\nu_1+\nu_2$ and $p_1+p_2$, respectively. Furthermore, if a theory has $\nu=0$, then partition functions do not have any phase ambiguity and hence we must have $p=0$.
From these properties, we conclude that there must be a homomorphism 
$
\bZ_{16} \ni \nu \mapsto p \in \bR/\bZ.
$
We will show in Sec.~\ref{sec:free} that this homomorphism is given by
\begin{equation}
\dimreduce_\CC: \bZ_{16}\ni 1 \mapsto \frac{1}{16} \in \bR/\bZ.\label{x}
\end{equation}

\subsection{Anomalies and projective representations}
More general treatment is as follows. (The reader can skip this subsection on a first reading and go to Sec.~\ref{sec:free}.)
As a warm-up, let us consider $G$-symmetric 0+1d systems. If the $G$ symmetry does not have an anomaly, the group $G$ acts on the Hilbert space $\cH$. 
If the $G$ symmetry is anomalous, the general principle says \cite{Chen:2011pg} that the anomaly is encoded by the 1+1d $G$-SPT phase, which is specified by a cohomology class $u\in H^2(BG,\U(1))$ as a Dijkgraaf-Witten theory \cite{Dijkgraaf:1989pz}.

When $u$ is nonzero, the group $G$ acts on the Hilbert space $\cH$ projectively, or equivalently, a nontrivial central extension $\hat G$ \begin{equation}
0\to \U(1)\to \hat G\to G \to 0
\end{equation} acts linearly on $\cH$. 
It is a standard mathematical fact that such central extensions are classified by the same cohomology group $ H^2(BG,\U(1))$. 
So, the same cohomology class $u$ specifies both the  $G$-SPT phase in the 1+1d bulk 
and the class of the projective $G$ representation on the 0+1d boundary.

As a second warm-up, consider a $G$-symmetric 1+1d system with an anomaly characterized by an element $u\in H^3(BG,\U(1))$. Put such a system on a spatial circle $S^1$ with the holonomy $g\in G$. The corresponding Hilbert space $\cH_g$ carries a projective representation of $C_g(G)$, the centralizer of $g$ in $G$, whose class as a projective representation is given by a certain class $\dimreduce_{S^1_g}(u) \in H^2(BC_g(G),\U(1))$, where \begin{equation}
\dimreduce_{S^1_g} : H^3(BG,\U(1)) \to H^2(BC_g(G),\U(1))
\end{equation} is a certain homomorphism whose explicit form is given e.g.~in \cite{Dijkgraaf:1989pz}.
From the point of view of the bulk SPT, this map $\dimreduce_{S^1_g}$  specifies the 1+1d $C_g(G)$-SPT resulting from the $S^1$ compactification with holonomy $g$ of the 2+1d $G$-SPT. 

We are interested in the time-reversal anomaly of 2+1d  systems with fermions such that $\time^2=(-1)^F$. In the following, we will call such systems 2+1d pin$^+$ systems, since fermions with $\time^2=(-1)^F$ correspond to having a pin$^+$ structure on non-orientable manifolds \cite{Kapustin:2014dxa,Witten:2015aba}.
As argued there, the corresponding 3+1d fermionic SPT phase is characterized by the dual of the bordism group given by $\Omega^4_\text{pin$^+$}=\bZ_{16}$.
Let us choose the spatial slice to be the crosscap geometry \eqref{ccgeometry} with a chosen pin$^+$ structure (which will be discussed more explicitly in the next subsection). The geometry has a $\U(1)$ isometry, and on an anomalous system it can be realized projectively. The class $\vartheta\in H^2(B\U(1),\U(1))=\bR/\bZ$ specifying the class of the projective representation is exactly the momentum mod 1; as a $\U(1)$-SPT phase in 1+1 dimensions, the parameter $\vartheta$ specifies the theta angle of the background $\U(1)$ gauge field.
Correspondingly, there should be a homomorphism \begin{equation}
\dimreduce_\CC: \Omega^4_\text{pin$^+$} \to H^2(B\U(1),\U(1)). 
\end{equation} 
This homomorphism can be determined by studying on the crosscap geometry a system whose time-reversal anomaly is known.  We will find below that
it is given by \eqref{x}.

\subsection{Analysis of the free fermion system}\label{sec:free}
To see that the anomalous momentum is given by \eqref{x}, take the 2+1d time-reversal invariant massless Majorana fermion system, 
which has the anomaly $1\in \bZ_{16}$ as computed in \cite{Witten:2015aba}. We need to compute its momentum on the crosscap geometry. 
This computation can be done by borrowing the results of \cite{Hsieh:2015xaa} where the background momentum on the Klein bottle was essentially computed. 
Here we give a simplified version of their arguments.

The geometry of the Klein bottle is given by \begin{equation}
\begin{aligned}
(x,\theta)&\sim (-x,\theta+\pi), \\
(x,\theta)&\sim (x+2L,\theta).
\end{aligned}
\label{kbgeometry}
\end{equation}
We have two crosscaps at $x=0$ and $x=L$. 

There are four possible pin$^+$ structures on the Klein bottle, as we can see as follows. First, under the identification $(x,\theta) \sim (-x,\theta+\pi)$, we can impose two possible conditions on the fermion $\psi$ as
\beq
\psi(x,\theta) = \pm \gamma_x \psi( - x, \theta+\pi), \label{eq:pinst1}
\eeq
where $\gamma_x$ is the gamma matrix in the $x$ direction which satisfies $(\gamma_x)^2=1$.\footnote{If we consider a $\pin^-$ fermion,
then the $\gamma_x$ in \eqref{eq:pinst1} is replaced by $i \gamma_x$ which satisfies $(i \gamma_x)^2=-1$.}
The choice of the $\pm$ sign in \eqref{eq:pinst1} represents the choice of the $\pin^+$ structure at the crosscap at $x=0$.
In the same way, we have another two possible $\pin^+$ structures at the crosscap at $x=L$ given by
\beq
\psi(x,\theta) = \pm \gamma_x \psi( 2L- x, \theta+\pi).\label{eq:pinst2}
\eeq

Some consequences of these $\pin^+$ structures are as follows. Under $\theta \sim \theta + 2\pi$, we always have periodic (R) boundary condition
\beq
\psi(x,\theta) = \psi(x,\theta + 2\pi), \label{eq:pinst3}
\eeq
which is a consequence of $(\pm \gamma_x)^2=1$.
In contrast, the boundary condition under $x \sim x+2L$ is given by
\beq
\psi(x,\theta) =(\pm 1) (\pm 1)\psi(x+2L,\theta)  \label{eq:pinst4}
\eeq
where the first and the second $(\pm 1)$ represent the signs in \eqref{eq:pinst1} and \eqref{eq:pinst2}, respectively.

We assume that each crosscap has its own anomalous momentum. 
The anomalous momentum of the crosscap at $x=0$ ($x=L$) with the $\pin^+$ structure \eqref{eq:pinst1} (\eqref{eq:pinst2})
are denoted as $p_\pm \in \bR$, where the subscript $\pm$ corresponds to the $\pin^+$ structures.
We remark that we consider these momenta as taking values in $\bR$ instead of $\bR/\bZ$ in this subsection.
This is necessary as we will see below.

When the spin structure along the $x$ direction is antiperiodic, meaning $(\pm 1) (\pm 1) = -1$ in \eqref{eq:pinst4},
there are no fermionic zero modes in the background, and there is no background momentum. 
Therefore, $p_+ + p_-=0$.

When the spin structure along the $x$ direction is periodic, meaning $(\pm 1) (\pm 1) = 1$, we consider
a Kaluza-Klein reduction in the $x$ direction. The system reduces to the massless 1+1d Majorana fermion system along $\theta$ with the periodicity $\theta\sim\theta+\pi$, up to massive Kaluza-Klein modes which do not contribute to the vacuum momentum.  
The conditions \eqref{eq:pinst1} and \eqref{eq:pinst2} mean that after the reduction, the 1+1d Majorana fermion is in the R-NS sector along $\theta \sim \theta+\pi$, 
where the R and NS sectors correspond to the components of $\psi$ which are the eigenvectors of $\pm \gamma_x$ with the eigenvalues $+ 1$ (for R) and $-1$ (for NS), 
respectively. If we choose the $+$ sign in \eqref{eq:pinst1} and \eqref{eq:pinst2}, the left-moving sector is periodic while the right-moving sector is antiperiodic. 
On the other hand, if we choose the $-$ sign, then the right-moving sector is periodic while the left-moving sector is antiperiodic.
Let us consider the case of the $+$ sign. On the $S^1$ given by $\theta \sim \theta + \pi$, the momentum 
of the vacuum of the R-NS sector which is appropriately normalized with respect to $\theta'=2\theta$  is given by
$1/24 - (-1/48)=1/16$. 
In the crosscap geometry \eqref{ccgeometry} the periodicity is actually $\theta\sim\theta+2\pi$. 
Therefore the momentum of the R-NS vacuum (normalized with respect to $\theta$) counts as the fractional momentum $1/8$ of the Klein bottle geometry. Therefore, $2p_+=1/8$.
We thus conclude that $p_\pm=\pm 1/16$. This is the relation \eqref{x} we wanted to show.

In the last step of the above discussion, we needed to divide the vacuum momentum by $2$ to go from $2p_+=1/8$ to $p_+=1/16$. 
We emphasize that this is possible
because we have treated the momentum as taking values in $\bR$ rather than $\bR/\bZ$. Otherwise, the division by 2 is not justified in 
$\bR/\bZ$. This is the reason that $\nu=8$ was not concluded to have an anomaly in \cite{Hsieh:2015xaa}.
By considering the vacuum momentum as taking values in $\bR$, we can see that $\nu=8$ has the anomalous momentum $p=1/2$.

\section{Time-reversal anomaly of topological theories}\label{how}
In this section we will explain how we can determine the time-reversal anomaly of 2+1d topological pin$^+$ theories.
In the following, we assume that the 1+1d RCFT corresponding to the 2+1d theory under consideration has the relation between
left and right central charges as $c_L=c_R$ so that the 2+1d theory has no framing anomaly, since 
we can choose no framing on non-orientable manifolds.\footnote{More precisely, the condition $c_L=c_R$ is derived as follows.
On oriented manifolds, we can always eliminate the framing anomaly (i.e., the dependence of the partition function on the trivialization of the tangent bundle)
at the cost of making the partition function depend on the metric through the $\eta$-invariant; see Sec.~2 of \cite{Witten:1988hf}. 
The dependence on the $\eta$-invariant is schematically given by $(c_L - c_R) \eta$. 
However, the $\eta$-invariant changes the sign under the change of the orientation of the manifold. Any theory with time-reversal symmetry must not depend on the choice of
orientation, and hence we must have $c_L=c_R$. 
However, we need to note that this condition is derived under the assumption that the 3+1d bulk contribution is absent.
If there are bulk terms such as $\pi \hat{A}$, we interpret them as a 2+1d  invertible field theory such as spin-Ising TQFT (for $\pi \hat{A}$)
or $\U(1)_{-1}$ Chern-Simons theory (for $2\pi \hat{A}$) and then we get $c_L=c_R$. This re-interpretation of the 3+1d bulk contributions as the 2+1d boundary theories
on oriented manifolds is possible because of the Atiyah-Patodi-Singer index theorem and the fact that the index $J$ of the Dirac operator coupled only to the metric satisfies $(-1)^J=1$ .
If the bulk contribution is $\frac{1}{2}\pi \hat{A}$ (which happens for $\nu=1$ mod 2) it is not possible to re-interpret the bulk term as a boundary theory
and the combined bulk-boundary system should be considered seriously even on oriented manifolds.  \label{longfootnote}
} 

To understand how we can study the fractional momentum carried by  the crosscap in these theories, it is useful to first recall the following fact.
A single quasiparticle $p$ (or, equivalently, a type of the line operator $p$) in a topological theory carries a spin $h_p$ mod 1.  If this quasiparticle $p$ is placed at the tip of the cigar, this spin translates to the anomalous momentum $h_p$ corresponding to the isometry of the cigar. Correspondingly, if we create a state $\ket{p}$ in the Hilbert space of the theory on $T^2=S^1_A\times S^1_B$ using the geometry of a disk times a circle, $D^2_A \times S^1_B$, with the line operator $p$ at the center of $D^2_A$ extending along $S^1_B$, it transforms under the transformation $T\in \SL(2,\bZ)$ as \begin{equation}
T: \ket{p}\mapsto e^{2\pi i h_p} \ket{p},
\end{equation} since $T$ changes the framing of the line operator by a single unit.\footnote{The reader should not confuse $T$ which is an element of $\SL(2,\bZ)$ and $\time$ which is the time-reversal.}

Therefore, to determine the time-reversal anomaly of a 2+1d topological pin$^+$ theory, we need to determine the $T$ eigenvalue of the crosscap state $\ket{\CC }$ on $T^2$ created by the geometry $\MO  _A\times S^1_B$, where $\MO  _A$ is the M\"obius strip, connecting the boundary $S^1_A$ and the crosscap
\beq
\MO  _A=\{ (x,\theta) \in [-1, 1] \times \bR ; (x,\theta) \sim (-x,\theta+\pi) \}  .
\eeq 
The boundary $S^1_A = \partial \MO  _A$ is given by $(x=1, \theta)$ with $\theta \sim \theta+2\pi$, and the crosscap is at $(x=0,\theta)$ with $\theta \sim \theta+\pi$.
We note here that the spin structure around $S^1_A$ is necessarily periodic, since this direction wraps the crosscap twice, while we are considering a pin$^+$ theory;
see \eqref{eq:pinst3}.
If we were considering a pin$^-$ theory, the spin structure around $S^1_A$ would be antiperiodic instead.
The spin structure around $S^1_B$, in contrast, can be chosen at will. 
In the following, we will always take it to be antiperiodic, to be specific.

The state $\ket{\CC}$ must be an eigenstate of $T$,
\beq
T \ket{\CC} = e^{2\pi i p} \ket{\CC}. \label{eq:Teigenstate}
\eeq
The reason is that the action of the Dehn twist $T$ does not change the topology of the geometry $\MO  _A\times S^1_B$,
and hence the physical states before and after the action of $T$ must be the same in a topological theory. 
Physical states correspond to rays in the Hilbert space,
and hence $ \ket{\CC}$ and $T \ket{\CC}$ must be proportional to each other. Furthermore, this eigenvalue is the exponential of the crosscap momentum as discussed above.

In a unitary 2+1d topological theory, we are given a collection of quasiparticles (i.e., types of line operators) equipped with the fusion products and other data. Among them, we have the standard conjugation $p\mapsto \bar{p}$ associated to the CPT (or more precisely CRT) transformation. In a time-reversal invariant theory we also have the time reversal  $p\mapsto \time p$. 
We prefer to use the spatial reflection $\reflection$, which is given by $p\mapsto \reflection p := \time\bar{p}.$
The conjugation does not change the spin $h_p$ mod 1, while the time reversal and the spatial reflection change the sign of the spin: $h_p \mapsto -h_p$.

From the geometry we can see that \begin{equation}
\reflection_A \ket{\CC }= \ket{\CC '}
\end{equation} where $\reflection_A$ is the reflection $\theta \mapsto -\theta$, 
and $\ket{\CC '}$ is the state created by the crosscap with the opposite pin$^+$ structure.
More precisely, we take the $+$ sign in \eqref{eq:pinst1} for $\ket{\CC }$ and the $-$ sign for $\ket{\CC' }$.
The reason for the change of the $\pin^+$ structure is that $\reflection_A$ acts as $\reflection_A(\psi)(x,\theta) = \gamma_\theta \psi(x,-\theta)$, and
the sign in \eqref{eq:pinst1} changes
 because of the anti-commutation $\gamma_x \gamma_\theta= - \gamma_\theta \gamma_x$.
In the same way, one can see that the reflection $\reflection_B$ in the $S^1_B$ direction also changes the $\pin^+$ structure.

Under the above transformations, the $T$ eigenvalues of $\ket{\CC }$ and $\ket{\CC '}$ should be inverse to each other.
In a non-spin theory, there is no distinction of $\ket{\CC }$ and $\ket{\CC '}$. 
Therefore the only possible $T$ eigenvalues are $\pm1$, corresponding to $0,8$ mod $16$ in the $\bZ_{16}$ classification, as it should be.

In a spin topological theory, there is a distinguished quasiparticle $f$ whose corresponding loop operator measures the spin structure; it represents the transparent fermion.  
It is a c-number ``operator'' which can be constructed purely from the background metric~\cite{Seiberg:2016gmd}.
Because it has the spin $1/2$, it has a framing anomaly which corresponds to the choice of the spin structure of the tangent bundle of the loop $C$ of this operator.
By fixing the spin structure of the tangent bundle of $C$ to be anti-periodic, the spin structure of the normal bundle 
(which is measured by the value of $f$) is determined from the spin structure
of the underlying manifold.
Then $f$ takes the value $+1$ on a cycle with NS boundary condition and $-1$ on a cycle with R boundary condition.

The braiding of any line operator with $f$ is  either $+1$ or $-1$;
the former is the standard NS quasiparticles, and the latter is the R ``quasiparticles''\footnote{They do not correspond to any dynamical excitations and instead they change
the background geometry.
In that sense they may be called more properly as R line defects.} around which we have R spin structure. 

As discussed above, the M\"obius strip $\MO  _A$ automatically has the periodic spin structure around the boundary circle $S_A$. Therefore, we should be able to expand $\ket{\CC }$ as \begin{equation}
\ket{\CC }=\sum_{p: \text{R quasiparticle}} c_p  \ket{p},
\end{equation}

Given a quasiparticle $p$, denote the corresponding loop operator wrapping $S^1_A$ or $S^1_B$ of the 
$T^2 = \partial (\MO  _A\times S^1_B)$ by $A(p)$ and $B(p)$, respectively. For an NS line $p$, the crosscap state should satisfy
\begin{equation}
B(p)\ket{\CC }=B(\reflection p)\ket{\CC }\label{Brelation}
\end{equation} since we can move the quasiparticle $p$ across the crosscap to make it to be $\reflection p$.%
\footnote{The \eqref{Brelation} is valid when $p$ is an NS line. For an R line,
one can check that the spin structure of \eqref{eq:pinst1} is changed when we move the R line, and hence the complementary state $\ket{\CC'}$ appears
as $B(p)\ket{\CC }=B(\reflection p)\ket{\CC'}$.
We also remark that the two states 
$A(p)\ket{\CC }$ and  $A(\reflection p)\ket{\CC }$ are also related,
but the precise relation depends on the braiding of $p$ and $\reflection p$, since to move the line of $p$ wrapped around the boundary of the M\"obius strip across the central crosscap, it needs to braid nontrivially with its self-reflection.
}

In the next section, to determine $\ket{\CC }$, we use the  
conditions discussed above, namely:
\begin{enumerate}
\item the fact that $\ket{\CC}$ is an eigenstate of $T$ as in \eqref{eq:Teigenstate}, $T \ket{\CC} = e^{2\pi i p} \ket{\CC}$,  
\item $A(f)\ket{\CC } = -\ket{\CC } $ and $B(f)\ket{\CC } = +\ket{\CC } $, corresponding to R and NS boundary conditions on $S_A$ and $S_B$, respectively,
and
\item the consistency of the action of the loop operator around $S^1_B$ \eqref{Brelation}, $B(p)\ket{\CC }=B(\reflection p)\ket{\CC }$.
\end{enumerate}

\section{Examples}\label{examples}
To illustrate the discussions so far, in this section we consider a few examples.

\subsection{Semion-fermion theory}\label{sec:semionfermion}
We first discuss the semion-fermion theory introduced in the condensed matter literature \cite{Fidkowski:2013jua}. As a Chern-Simons theory it is realized as $\U(1)_{2}\times \U(1)_{-1}$ and was discussed in \cite{Seiberg:2016rsg}. In appendix~\ref{app:A}, we show that this theory does arise on a boundary of  a topological superconductor of $\nu=\pm 2$ by using the general methods developed in \cite{Tachikawa:2016xvs}.

Let us denote the unique nontrivial line operator of $\U(1)_2$ by $s$, representing the semion.
The spin is $1/4$, and it satisfies $s^2=1$. 
We use the symbol $f$ for the spin $-1/2$ operator of $\U(1)_{-1}$, which is the transparent fermion of the spin topological theory. 
To describe the R-sector, we need another line $r$ of $\U(1)_{-1}$, whose spin is $-1/8$, with the property $r^2=f$. A convenient way to 
consider this R-sector line $r$ is to start from a non-spin $\U(1)_{-4}$ theory whose fundamental line is $r$, and then divide the gauge group $\U(1)$ by $\bZ_2$.
The gauge field $a'_\mu$ for $\U(1)_{-4}$ is related to the gauge field $a_\mu$ of $\U(1)_{-1}$ by $a_\mu = 2a'_\mu$,
and hence $r$ may be regarded as a loop operator with the half-integral charge $1/2$ of $\U(1)_{-1}$.

The consistency with the fact that $\reflection$ changes the sign of the spin $h_p$ requires that
$\reflection(s)=fs$.
The candidates for the crosscap state satisfying the conditions discussed in the previous section are \begin{equation}
\ket{\CC }_{{\rm SF}_-} \propto\ket{r}+\ket{r^3} \label{eq:sf-}
\end{equation} or \begin{equation}
\ket{\CC }_{{\rm SF}_+}  \propto\ket{sr}+\ket{sr^3}. \label{eq:sf+}
\end{equation}
The $T$ eigenvalues are $e^{-2\pi i/8}$ and $e^{+2\pi i/8}$, respectively.
With the former choice, the time-reversal anomaly is $\nu=-2$ while with the latter we have $\nu=+2$. 
We denote the theories with $\nu=+2$ and $\nu=-2$ as ${\rm SF}_+$ and ${\rm SF}_-$, respectively.

In general, given a TQFT on oriented manifolds, we need more detailed information about the action of time-reversal symmetry
to formulate it on non-orientable manifolds. This is analogous to the fact that two transformations of majorana fermions $\time(\psi)=+\gamma_0\psi$ and $\time(\psi)=-\gamma_0\psi$
correspond to two different values $\nu=+1$ and $\nu=-1$.
The above result suggests that there are two ways to couple the theory $\U(1)_{2}\times \U(1)_{-1}$
to the geometry of non-orientable manifolds, and they give the values $\nu=+2$ and $\nu=-2$, respectively. 

It is difficult to see this, however, from a more traditional point of view, 
because the time reversal $\time$ is realized as a quantum symmetry rather than the symmetry of the classical action. 
More precisely, the problem is that $\reflection$ must satisfy $\reflection^2=1$ in the $\Pin^+$ group,
but this relation is not realized at the classical level and only achieved at the quantum level 
 \cite{Seiberg:2016rsg}.
Therefore it is not straightforward to put the theory on non-orientable manifolds. 
We leave it a future work to study the full details.

\subsection{T-Pfaffian theory}\label{sec:tpfaffian}
\paragraph{Specification of the theory:}
The T-Pfaffian theory is the name given to the topological theory $(\U(1)_{-8}\times \text{Ising}_{1/2})/\bZ_2$  by the condensed-matter theorists. 
Here, Ising$_{1/2}$ is the non-spin Ising TQFT with the right-moving central charge $+1/2$,
and we take $\U(1)_{-8}$ to have the left-moving central charge $1$. 
To put the system on non-orientable manifolds, we need to cancel  the total central charge, as we discussed in footnote \ref{longfootnote}. 
For this purpose we need an almost trivial spin TQFT with the right moving central charge $+1/2$ such that there is only one state on any spatial slice.%
\footnote{\label{IFT} 
The theories with only one state on any spatial slice are called invertible field theories. 
Let ${\rm IFT}_{c}$ be the invertible field theory on oriented spin manifolds whose partition function is given by $\exp ( - c i \pi \eta)$, where $\eta$
is the Atiyah-Patodi-Singer eta invariant of a fermion coupled only to metric on 2+1d. The $c$ corresponds to the framing anomaly.
The smoothness of the partition function requires that $2c$ must be an integer for invertible field theories in the normalization of $\eta$ as in \cite{Seiberg:2016rsg}.
They have a property that ${\rm IFT}_{c} \times {\rm IFT}_{c'}={\rm IFT}_{c+c'}$ and in particular ${\rm IFT}_{c} \times {\rm IFT}_{-c}$ is the trivial theory.
For example, we can realize them as $\SO(n)_{1} = {\rm IFT}_{n/2}$, $\U(1)_{1} = {\rm IFT}_1$, sIsing$_{\pm1/2}=\text{IFT}_{\pm1/2}$ etc.
See also Appendix C.5 of \cite{Seiberg:2016rsg}.
}
The spin Ising TQFT sIsing$_{+1/2}$ does the required job.\footnote{For the detailed discussions of the relation between the non-spin Ising TQFT and the spin Ising TQFT, see \cite{Gaiotto:2015zta}.}
Therefore the T-Pfaffian theory we consider is \begin{equation}
[(\U(1)_{-8}\times \text{Ising}_{1/2})/\bZ_2] \times \text{sIsing}_{1/2}.
\end{equation}
The quasiparticles of $\U(1)_{-8}$ are denoted by $c^k$, whose spin is $-k^2/16$. 
We denote the quasiparticles of the Ising$_{1/2}$ and sIsing$_{1/2}$ theories by $\psi$, $\sigma$ and $\psi'$, $\sigma'$, with spins $1/2$, $1/16$ and $1/2$, $1/16$, respectively.

We use $f:=c^4 \psi$ to form the $\bZ_2$ quotient.  Therefore, $f$ and $\psi'$ are transparent fermions. We need to keep in mind that the non-anomalous $\bZ_2$ one-form symmetry generated by $F:=f\psi'$ is gauged \cite{Bhardwaj:2016clt}.\footnote{As argued in \cite{Bhardwaj:2016clt}, gauging a non-anomalous $\bZ_2$ one-form symmetry whose corresponding line operator is $F$ has two main effects: i) it projects out line operators that non-trivially braid with $F$, ii) any two line operators $p$, $q$ that satisfy $pF=q$ under the fusion product are identified, and iii) any line operator $p$ that satisfies $pF=p$ in the fusion product splits into two operators $p_+$ and $p_-$. In our case, the first effect just means that we always pair an NS line operator from $A:= (\U(1)_{-8}\times \text{Ising}_{1/2})/\bZ_2$ and an NS line operator $B:=\text{sIsing}_{-1/2}$, or an R line operator from $A$ and an R-line operator from $B$. The second effect identifies $f$ and $\psi'$. As for the third effect, there is no line operator that satisfies $p=pF$ in our theory, so it does not play a role. This third effect, however, becomes important e.g.~when we check the relation $\text{sIsing}_{1/2}\times \text{sIsing}_{1/2}=\U(1)_{1}$.}

\paragraph{List of quasiparticles:}
NS quasiparticles and their spins are the ones given below:  \begin{equation}
\begin{array}{c||c|c|c|c|c|c|c|c}
& 1 & c & c^2 & c^3 & c^4 & c^5 & c^6 & c^7 \\
\hline
\hline
1 & 0 &  & \frac34 &  & 0 &  &\frac34&  \\
\hline
\sigma &  & 0 &  & \frac12 &  & \frac12& & 0\\
\hline
\psi & \frac12 &  & \frac14 &  & \frac12 &  &\frac14&  \\
\end{array}.\label{NStable}
\end{equation}
Multiplying by $ \psi'$ does not give new quasiparticles, since it is equivalent to multiplying by $f=c^4\psi$. 

R quasiparticles and their spins are the ones given below: \begin{equation}
\begin{array}{c||c|c|c|c|c|c|c|c}
& 1 & c & c^2 & c^3 & c^4 & c^5 & c^6 & c^7 \\
\hline
\hline
1\sigma' &  & 0 &  & \frac12 &   & \frac12 &  & 0\\
\hline
\sigma  \sigma' & \frac18 &  & \frac78 &  &  \frac18 &  & \frac78 & \\
\hline
\psi\sigma' &  & \frac12 &  & 0 &   & 0 &  & \frac12
\end{array}.
\end{equation} 
Note that the two R-lines listed above related by multiplying by $c^4\psi$, are in fact identical, since $c^4\psi$ is identified with $\psi'$, and $\psi'\sigma'=\sigma'$.

The spatial reflection should reverse the spin mod 1 of the quasiparticles.
To match what condensed matter physicists discuss, the spatial reflection also needs to reverse the power of $c$, and to fix $\psi$, $\sigma$ and $\tilde\psi$, $\tilde\sigma$.
These conditions uniquely determine the spatial reflection.
For example, we have $c^{1+2k}\sigma \leftrightarrow c^{7-2k}\sigma$ and $c^{2k} \leftrightarrow c^{8-2k} \psi^k$ for integer $k$.

\paragraph{Crosscap states:}
We can easily find one crosscap state that satisfies the condition \eqref{Brelation}: \begin{equation}
\ket{\CC }_{\text{T-Pfaffian}_+}\propto \ket{c\sigma'} + \ket{c^3\psi \sigma'} + \ket{c^5\psi\sigma'} + \ket{c^7\sigma'} \propto\ket{c\sigma'}+\ket{c^7\sigma'},
 \label{eq:crosscapTP0}
\end{equation} whose $T$ eigenvalue is $e^{2\pi i 0/16}$. The time-reversal anomaly is then $\nu=0$.

In general, once we find a state $\ket{\CC }_X$ satisfying the condition \eqref{Brelation}, we can find other states satisfying at least the same condition  \eqref{Brelation} as
\beq
\ket{\CC}_{p X} := B(p) \ket{\CC}_X.\label{eq:trick}
\eeq
This is because any two operators $B(p)$ and $B(q)$ commute by a topological reason; we can exchange the positions of the lines $B(p)$ and $B(q)$
without crossing them with each other. Hence we have 
\begin{equation}
\begin{aligned}
B(q) \ket{\CC}_{p X}&  = B(q) B(p) \ket{\CC}_X  = B(p) B(q) \ket{\CC}_X \\
& = B(p) B(\reflection q) \ket{\CC}_X= B(\reflection q)  \ket{\CC}_{p X} . 
\end{aligned}
\end{equation}
However, the condition \eqref{eq:Teigenstate} that the state $\ket{\CC}_{p X}$ be an eigenvector of $T$ is
not necessarily satisfied for all $p$. 

In the case of the T-Pfaffian, one can check that \eqref{eq:Teigenstate} is satisfied if $p=1$ or $c^4$. (Incidentally, these two lines $p=1$ and $c^4$ form the $\bZ_2$ one-form symmetries of the TQFT.)
The  crosscap state for $p=c^4$ is
\begin{equation}
\ket{\CC}_{\text{T-Pfaffian}_-}\propto \ket{c\psi\sigma'} + \ket{c^3 \sigma'} + \ket{c^5\sigma'} + \ket{c^7\psi\sigma'} \propto\ket{c^3\sigma'}+\ket{c^5\sigma'}, \label{eq:crosscapTP8}
\end{equation}
whose $T$ eigenvalue is $e^{2\pi i 8/16}$, meaning that $\nu=8$.

To conclude this subsection, we found that the T-Pfaffian theory has two different variants on non-orientable manifolds. One choice has $\nu=0$ with the crosscap state \eqref{eq:crosscapTP0},
and another has $\nu=8$ with the crosscap state \eqref{eq:crosscapTP8}.
We call these variants T-Pfaffian$_+$ and T-Pfaffian$_-$.

\subsection{Theories obtained by gapping free fermions}\label{sec:TPfSF}
In \cite{Seiberg:2016rsg} Seiberg and Witten considered a weakly-coupled system of fermions, scalars and a $\U(1)$ gauge field  such that in one phase we have $\nu=2r$ Majorana fermions and in other phase we have certain TQFTs. 
After quickly reviewing their construction, 
we apply our methods to the resulting TQFTs and show that we can correctly reproduce the expected value of $\nu$.

\def\ssa{\mathsf{a}}
\def\sss{\mathsf{s}}
\def\ssc{\mathsf{c}}

\subsubsection{Quick review}
We start from $r$ complex fermions $\chi_i$, $i=1,\ldots,r$, all of charge $2$ under the $\U(1)$ gauge field $\ssa$.
We also introduce a complex scalar $w$ of charge 1 and another complex scalar $\phi$ of charge 4.
We include the Yukawa coupling $\bar \phi \chi_{ai}\chi_{bi}\epsilon^{ab} +\text{c.c}$ in the theory, where $a,b$ are spinor indices.  
This system is time-reversal invariant when we give appropriate transformation rules.
We regard the neutral combination $\chi_i \bar w^2$ to have the same quantum numbers as the bulk 3+1d fermion, so that it can escape to the bulk.

Depending on the potential of $w$ and $\phi$, we can either give a vev to $w$ or $\phi$. In the former case, $\U(1)_\ssa$ is completely broken by eating $w$, and we just have $\nu=2r$ Majorana fermions. In the latter case, the vev of $\phi$ breaks $\U(1)_\ssa$ to $\bZ_4$. This can be represented by a $\U(1)^2$ Chern-Simons theory by introducing an additional Lagrange-multiplier gauge field $\ssc$, with the action \begin{equation}
\frac{4}{2\pi} \ssc d\ssa.\label{eq:cda}
\end{equation} Further, the $r$ Dirac fermions become massive by the vev of $\phi$ and can be integrate out.
When $r$ is even, the integrating-out does not produce any terms.
When $r$ is odd, the integrating-out generates the term \begin{equation}
\frac{2}{4\pi} \ssa d\ssa\label{eq:ada}
\end{equation} in addition to the Ising TQFT sector. For more details, see  \cite{Seiberg:2016rsg}. There they considered a more general class of theories where $\chi$ has charge $2s$, and we set $s=1$ for simplicity.

\subsubsection{Even number of complex fermions}
\paragraph{Specification of the theory:} The $\bZ_4$ gauge theory described by \eqref{eq:cda} has total central charge zero, and has 16 line operators $a^{m} c^n :=e^{im\oint \ssa+in \oint \ssc}$, $(m,n=0,1,2,3)$, whose spin is $mn/4$. 
This is however not the whole story. This $\bZ_4$ gauge theory does not feel the spin structure, but  we started from the theory that depends on the spin structure.

Therefore, we also have an almost trivial spin TQFT with zero central charge, with transparent line operator $\psi$ of spin $1/2$. There are two types of R-sector lines $\rho$ and $\rho'$, both of spin 0, such that $\rho^2=\rho'{}^2=1$ while $\rho\rho'=\psi$.\footnote{An explicit example of the construction
of such an almost trivial spin TQFT is given by $\U(1)_1 \times \U(1)_{-1} = [ (\U(1)_4 \times \U(1)_{-4})/\bZ_2]/\bZ_2$. Let $s$ and $t$ be the 
basic line operators of $\U(1)_4$ and $\U(1)_{-4}$, respectively. Then the quotient in $(\U(1)_4 \times \U(1)_{-4})/\bZ_2$ 
is taken with respect to the line operator $s^2 t^2$ with spin 0. The result of this quotient is a TQFT which contains four
line operators; $1=s^2t^2$, $\rho :=st=s^3t^3$, $\rho':=st^3=s^3t$ with spin 0 and $\psi := s^2 =t^2$ with spin $1/2$.
The CRT acts as $\mathsf{CRT}(s^n t^m)=s^{-n} t^{-m}$.} 
They are all self-conjugate under the CRT.
In the quick review above, we said that the combination $\chi_i \bar w^2$ can escape to the bulk.
This means in the TQFT language that the transparent fermion line defining the spin TQFT is not $\psi$ but $f:=\psi a^2$.

\paragraph{List of quasiparticles:} The NS-sector lines are then \begin{equation}
\begin{array}{rllll}
\text{even $n$}:& a^m c^n& (\text{spin  $mn/4$}), &a^m c^n \psi &(\text{spin $mn/4+1/2$}), \\
\text{odd $n$}:& a^m c^n\rho &(\text{spin  $mn/4$}),& a^m c^n \rho' &(\text{spin $mn/4$}) 
\end{array}
\end{equation}
whereas
the R-sector lines are then \begin{equation}
\begin{array}{rllll}
\text{odd $n$}:& a^m c^n& (\text{spin  $mn/4$}), &a^m c^n \psi &(\text{spin $mn/4+1/2$}), \\
\text{even $n$}:& a^m c^n\rho &(\text{spin  $mn/4$}),& a^m c^n \rho' &(\text{spin $mn/4$}) .
\end{array}
\end{equation}

The exponent of $a$ is the electric charge, and that of $c$ is the vorticity. As such, under the time reversal, the former is reversed while the latter is kept.  In particular, $c\rho$ is mapped by $\time$ to either $c\rho$ or $c\rho'$.  Which is the case can be determined from the high-energy realization. If we start from $r$ Dirac fermions, there are $r$ fermionic zero-modes at the core of the vorticity-one vortex, forming spinor representations of $\SO(r)$. Here $r$ is even, and therefore they split into two chiral spinors, and they correspond to $c\rho$ and $c\rho'$. The time-reversal acts by complex conjugation. Therefore, when $r=0$ mod 4, $c\rho$ is mapped to $c\rho$, while when $r=2$ mod 4, $c\rho$ is mapped to $c\rho'$.
Correspondingly, under the spatial reflection $\reflection$, 
$c\rho$ is mapped to $c^3\rho$ or $c^3\rho'$ depending on whether $r=0$ mod 4 or $r=2$ mod 4.

\paragraph{Crosscap states:} 
Using the data determined above, we can find the following four crosscap states: \begin{equation}
\begin{array}{r@{\,}l@{}l@{}l@{}l}
\ket{\CC}_{r=0}&\propto  \ket{c} &+ \ket {c^3}& + \ket{a^2c\psi }&+\ket{a^2 c^3\psi},\\
\ket{\CC}_{r=2}&\propto \ket{ac} &+ \ket {ac^3\psi} &+ \ket{a^3c\psi }&+\ket{a^3 c^3},\\
\ket{\CC}_{r=4}&\propto \ket{a^2c} &+ \ket {a^2c^3} &+ \ket{c\psi }&+\ket{ c^3\psi}, \\
\ket{\CC}_{r=6}& \propto\ket{a^3c}& + \ket {a^3c^3\psi} &+ \ket{ac\psi }&+\ket{a c^3}.
\end{array}
\end{equation}
They satisfy all the conditions discussed in the previous sections, and has the correct eigenvalue $e^{2\pi i (2r)/16}$ under $T\in \SL(2,\bZ)$.
This is consistent with the identifications $\nu=2r$.

\subsubsection{Odd number of complex fermions}\label{sec:oddr}
\paragraph{Specification of the theory:}
The Chern-Simons sector has the action given by the sum of \eqref{eq:cda} and \eqref{eq:ada}.
We can diagonalize the kinetic term by setting $\sss=\ssa+2\ssc$: \begin{equation}
\frac{1}{4\pi}(-8\ssc d\ssc + 2 \sss d\sss).
\end{equation}
We denote the Wilson line operators by $s^{m} c^n :=e^{im\oint \sss+in \oint \ssc}$, $(m=0,1; n=0,1,\ldots, 7)$.
In addition, we have an Ising sector Ising$_{1/2}$ of left-moving central charge $+1/2$,
with the line operators  $\psi$, $\sigma$ of dimension $1/2$, $1/16$ respectively.
The transparent fermion corresponds to the operator $f:=c^4 \psi$, with respect to which we take the $\bZ_2$ quotient.
Notice that $s^2=1$ implies $c^4=a^2$ and hence we can also write $f=\psi a^2$ as in the case of even $r$.

So far, we have the topological theory 
$\U(1)_2 \times (\U(1)_{-8}\times \text{Ising}_{1/2})/\bZ_2$.
This has an uncancelled total central charge $+1/2$.  
We then need to multiply it by a trivial spin theory of central charge $-1/2$, which is given by the spin Ising theory sIsing$_{-1/2}$ with the line operators $\tilde\psi$, $\tilde\sigma$ of dimension $-1/2$, $-1/16$ respectively.
The final theory is \begin{equation}
\U(1)_2 \times (\U(1)_{-8}\times \text{Ising}_{1/2})/\bZ_2 \times \text{sIsing}_{-1/2}\label{finaltheory}
\end{equation}
and the theory manifestly free of the framing anomaly.\footnote{The final spin Ising part was implicit in \cite{Seiberg:2016rsg}, and was represented using the bulk $\hat A$ genus there.
See the discussion of the footnote~\ref{longfootnote}.}
This is the theory discussed in Sec.~6 of \cite{Seiberg:2016rsg} and Sec.~3.2.3 of \cite{Witten:2016cio}.  

\paragraph{Factorization of the theory:}
The structure of \eqref{finaltheory} is consistent with the factorization of the theory as T-Pfaffian $\times$ semion-fermion. Recall that 
\beq
\text{semion-fermion}:~& \U(1)_2 \times \U(1)_{-1}, \\
\text{T-Pfaffian}:~
& (\U(1)_{-8}\times \text{Ising}_{1/2})/\bZ_2 \times \text{sIsing}_{+1/2} .
\eeq 
Their product can be simplified using the multiplication rule of the invertible field theories $\text{IFT}_c$ given in footnote \ref{IFT}, and the result reproduces the theory \eqref{finaltheory}.
We studied the semion-fermion  in Sec.~\ref{sec:semionfermion} and we saw there that $\nu=\pm 2$;
the T-Pfaffian was studied in Sec.~\ref{sec:tpfaffian} and gave $\nu=0$ or $\nu=8$.
There are four ways to combine them. 

Let $X$ be the theory 
\beq
X = \text{SF}_- \times \text{T-Pfaffian}_+
\eeq
where the crosscap states of $\text{SF}_- $ and $ \text{T-Pfaffian}_+$ are given in \eqref{eq:sf-} and \eqref{eq:crosscapTP0} respectively.
In this product, the transparent fermions of $\text{SF}_- $ and $ \text{T-Pfaffian}_+$ are identified.
The $X$ has the crosscap state $\ket{\CC }_{X}=\ket{\CC }_{\text{SF}_-} \otimes \ket{\CC }_{\text{T-Pfaffian}_+}$.
Also let $pX$ be the theory whose crosscap state is given as $\ket{\CC}_{pX}:=B(p)\ket{\CC}_X$ for $p=1,s, c^4,$ and $s c^4$.
More explicitly, $sX= \text{SF}_+ \times \text{T-Pfaffian}_+$, $c^4 X = \text{SF}_- \times \text{T-Pfaffian}_-$ and 
$sc^4 X = \text{SF}_+ \times \text{T-Pfaffian}_-$.

The time-reversal anomalies of theories with odd $r$ is summarized in the following table:  \begin{equation}
\begin{array}{c|cccc}
\nu& 2 & 6 & 10 & 14 \\
\hline
\text{theory} & sX & c^4 X & sc^4 X & X
\end{array}. \label{eq:possiblenu}
\end{equation}
Thus we can consistently make identifications $\nu=2r$.

\subsection{Speculations on $\time^2$ of quasiparticles}

In the above analyses, we have obtained the values of $\nu$ in various theories by finding crosscap states satisfying the consistency conditions discussed in Sec.~\ref{how}.
The set of values of $\nu$ obtained in that way perfectly matches the ones found in \cite{Metlitski:2014xqa,Seiberg:2016rsg,Witten:2016cio}.
However, the following point needs to be noticed. In \cite{Metlitski:2014xqa,Seiberg:2016rsg,Witten:2016cio},
 the distinction between different values of $\nu$ was to be found in the eigenvalues of the square of the time-reversal operation $\time^2$ acting on various quasiparticles, 
 but our discussion has not used this information yet. There should be a general way to find the correspondence
 between $\time^2$ eigenvalues and crosscap states.

We remark that what we are discussing here is not the change of the types of quasiparticles under $p \to \time p \to \time^2 p=p$,
but the eigenvalues of $\time^2 $ which, in the language of the low energy TQFT, might be given by the action of $\time^2$ on the Hilbert space
on a spatial slice with a time-like Wilson line of a quasiparticle $p$.\footnote{However, in a compact space without boundary, the Hilbert space with a single time-like Wilson line
is zero. It is necessary to find a proper definition of ``the eigenvalues of $\time^2$'' in the context of TQFT.}
In the UV description, it is an action of $\time^2$ on the states with the actual physical excitations corresponding to $p$.

The assignments of $\time^2$ are as follows, according to \cite{Metlitski:2014xqa,Seiberg:2016rsg,Witten:2016cio}, in our notation.
Let us consider semion-fermion and T-Pfaffian. 
According to the papers cited above, there is actually two versions of each of these theories, which we denote as SF'$_\pm$ and T-Pfaffian'$_\pm$.
The theories SF'$_\pm$ are characterized by the $\time^2$ eigenvalue acting on $s$ as
\begin{equation}
\time^2=\begin{cases}
+i & : \text{SF'}_+,\\
-i & : \text{SF'}_-
\end{cases}
\end{equation}
Similarly, the theories T-Pfaffian'$_\pm$ are characterized by the $\time^2$ eigenvalue acting on $c \sigma$ as
\begin{equation}
\time^2=\begin{cases}
+1 & : \text{T-Pfaffian'}_+,\\
-1 & : \text{T-Pfaffian'}_-
\end{cases}
\end{equation}
Then, all the results of this paper are consistent with the identification that $\text{SF'}_\pm=\text{SF}_\pm$ and $\text{T-Pfaffian'}_\pm=\text{T-Pfaffian}_\pm$,
where
\beq
\text{SF}_+:~\nu=2,~~~\text{SF}_-:~\nu=-2;~~~\text{T-Pfaffian}_+:~\nu=0,~~~\text{T-Pfaffian}_-:~\nu=8.
\eeq
For the theories studied in Sec.~\ref{sec:oddr} for odd $r$, the $\time^2$ eigenvalues are the ones obtained from 
the factorization $\text{SF}_\pm \times \text{T-Pfaffian}_\pm$.

If we have a theory $Y$, we get another theory $pY$ as $\ket{\CC}_{pY}=B(p)\ket{\CC}_Y$ for some $p$.
For example, $s  \text{SF}_+ = \text{SF}_-$ and $c^4  \text{T-Pfaffian}_+ = \text{T-Pfaffian}_-$.
Then, notice that we have the following braiding phases: \begin{equation}
\begin{array}{c|cc}
& c^4 & s \\
\hline
c\sigma &  -1 & +1 \\
s &  +1 & -1 
\end{array}.
\end{equation}
From these braiding, we find the following relationship: \begin{multline}
(\text{$\time^2$ of quasiparticle $q$ in theory $pY$})
=\\
(\text{braiding phase of $q$ and $p$})(\text{$\time^2$ of quasiparticle $q$ in theory $Y$}).
\label{eq:foo}
\end{multline} 
In our case, the theory $Y$ is $\text{SF}_\pm$, $\text{T-Pfaffian}_\pm$ or $\text{SF}_\pm \times \text{T-Pfaffian}_\pm$ ,
$p$ is either $c^4$ or $s$, and $q$ is either $c\sigma$ or $s$,
but the relation \eqref{eq:foo} seems general. 

The authors do not have a proper understanding of the relation \eqref{eq:foo}, mainly because they do not understand how the eigenvalues of $\time^2$ of quasiparticles are reflected in the language of TQFT. 
But the following argument seems to come close.

Consider the geometry $\MO_A\times S^1_B$ where $\MO_A$ is 
a M\"obius strip connecting a circle $S^1_A$ and a crosscap,
and consider a line $A(q)$ of the quasiparticle $q$ wrapping $S^1_A$.
This line of quasiparticle $q$ experiences the same parity flip twice, since the $A$-cycle wraps the crosscap twice.

Now, the difference between theories $Y$ and $pY$ might have an interpretation that the crosscap at the bottom of $\MO_A$ carries an additional insertion of a line $B(p)$ of quasiparticle $p$ along $S^1_B$.
Therefore, the way $A(q)$ acts is modified by a braiding of $A(q)$ with $B(p)$.
This seems to correspond to the braiding phase appearing in \eqref{eq:foo}.

\section*{Acknowledgements}
The authors thank Nati Seiberg and Edward Witten for posing the question for the authors to solve during private communications and helpful comments on the 
earlier version of the draft, 
and Edward Witten for helpful suggestions.
The authors also thank Lakshya Bhardwaj for discussions.

The work of Y.T. is partially supported in part by JSPS Grant-in-Aid for Scientific Research No. 25870159.
The work of K.Y. and Y.T. is supported by World Premier International Research Center Initiative
(WPI Initiative), MEXT, Japan.

\appendix

\section{Semion-fermion theory on a SYM domain wall} \label{app:A}
Here we show that the semion-fermion theory realized as $\U(1)_2 \times \U(1)_{-1}$ corresponds to $\nu=\pm 2$
by using the results of \cite{Tachikawa:2016xvs} concerning the domain wall of gauge theories.\footnote{The authors would like to thank Edward Witten whose suggestion led to this appendix.}
We also discuss certain generalizations, some of which may give gapped boundary theories of topological superconductors for odd $\nu$.

Let us consider a 3+1d ${\cal N}{=}1$ pure Super-Yang-Mills (SYM) theory with the gauge group $G$.
This is just a gauge theory with a minimally-coupled Majorana fermion $\lambda$ in the adjoint representation of $G$; this automatically leads to supersymmetry.
We assume that the gauge group is simple, connected and simply connected, $\pi_0(G)=\pi_1(G)=0$,
and the dual coxeter number $h^\vee$ is even, $h^\vee \in 2\bZ$. Also, the theta angle is assumed to be zero.
This theory confines and fermion condensation occurs with
\beq
\vev{\lambda\lambda}_k = \Lambda^3 e^{2 \pi i k/ h^\vee} \qquad (k=0,1,2,\cdots,h^\vee-1),
\eeq
where $\Lambda$ is the dynamical scale which can be assumed to be real and positive because the theta angle is zero.
There are $h^\vee$ vacua labelled by $k$.

We introduce a small real mass $m \lambda \lambda~(m \in \bR)$  for the majorana fermion $\lambda$. Then, the vacuum for $m>0$
is realized by the vacuum $k=0$ given as $\vev{\lambda\lambda}_{k=0}=\Lambda^3$, and the vacuum for $m<0$ is realized by the vacuum $k=h^\vee/2$ given as
$\vev{\lambda\lambda}_{k=h^\vee/2}=-\Lambda^3$. If we change the mass from positive to negative along one of the spatial directions (say $y=x^3$),
we get a domain wall interpolating them. Assuming that the time reversal symmetry is not spontaneously broken by the domain wall configuration,
a 2+1d boundary theory of the topological superconductor corresponding to $\nu=\pm \dim G$ is realized on this domain wall \cite{Tachikawa:2016xvs},
because $\lambda$ is in the adjoint representation which has dimension $\dim G$. The $\pm$ sign is determined by how the time reversal $\time$ acts on $\lambda$,
and for definiteness, we take it such that $\nu= \dim G$.

The domain wall exists even in the massless limit $m \to 0$ and the supersymmetry is restored in this limit.
Then, there is one massless goldstino on the domain wall associated to the spontaneous breaking of (super)translation invariance.
This fermion remains massless even if we introduce supersymmetry breaking mass $m$ because it is protected by the time reversal $\time$.
It is reasonable to assume that the goldstino provides the only massless fermionic degrees of freedom on the domain wall if the gauge group is simple.
Assuming that this is the case, the rest of the anomaly corresponding to $\nu'=\nu-1=\dim G-1$ is accounted for 
by the TQFT living on the domain wall.\footnote{It requires some computation to determine that the goldstino corresponds to $\nu=1$ rather than $\nu=-1$. 
} In fact, it was argued that some TQFT does live on the domain wall \cite{Acharya:2001dz,Gaiotto:2013gwa,Dierigl:2014xta,Gaiotto:2014kfa}.
Even without the time-reversal symmetry, the existence of some TQFT is required by the anomaly matching of the one-form global symmetry for $C(G)$,
where $C(G)$ is the center of the gauge group $G$~\cite{Gaiotto:2014kfa}.

Now let us focus our attention to the case  $G=\SU(2N)$ which has $h^\vee=2N \in 2\bZ$ and $\dim G=4N^2-1$.
The domain wall we are concerned with connects the vacuum $k=0$ and the vacuum $k=N$.
In this case, it was argued that there is a $\U(N)_{2N}$ Chern-Simons theory on the domain wall\footnote{Here we follow the convention common in the domain wall of supersymmetric theories. In the TQFT language, this corresponds to $(\U(1)_{2N^2} \times \SU(N)_{N})/{\bZ_N}$. }.
This theory should account for the anomaly $\nu'=\nu-1=4N^2-2$ of the time reversal symmetry.

Let $c$ be the framing anomaly (i.e.~the central charge of the corresponding RCFT) of this Chern-Simons theory.
Then we also need to introduce 3+1d bulk gravitational term $ 2\pi c \hat{A}$ to make the theory time-reversal invariant. 
A consistency check is that we must have the relation $4c = \nu' \mod 2$ which is required on orientable manifolds.
Indeed, $\nu'=4N^2-2$ and $c =1 + \frac{N}{2N} (N^2-1)$, so the condition is satisfied.

For example, the simplest case is given by the gauge group $G=\SU(2)$.
In this case, we have $N=1$ and $c=1$, and the bulk contribution $2\pi c \hat{A}$ may be replaced by a boundary invertible field theory with $c=-1$ (see the footnote~\ref{longfootnote}),
which we can take to be $\U(1)_{-1}$. Therefore, the total system is $\U(1)_2 \times \U(1)_{-1}$, at least on orientable manifolds. This is exactly the semion-fermion theory
discussed in Sec.~\ref{sec:semionfermion}. By the above construction, we have determined that this theory corresponds to $\nu'=2$ (or $\nu'=-2$ depending on the action of $\time$),
which perfectly agrees with the result of Sec.~\ref{sec:semionfermion}.

For $G=\SU(2N)$, the total system is $\U(N)_{2N} \times {\rm IFT}_{-c}$, where ${\rm IFT}_{-c}$ is an invertible field theory accounting for the framing anomaly $-c$.
Therefore, we conclude that it should be somehow possible to formulate the theory $\U(N)_{2N} \times {\rm IFT}_{-c}$ on non-orientable manifolds so that it reproduces 
the anomaly $\nu'=\pm (4N^2-2)=\pm 2 \mod 16$.

Finally, let us make a speculative comment. Under the above assumptions that (i) $\time$ is not spontaneously broken by the domain wall, and 
(ii) there is only one massless fermion on the domain wall which is the goldstino, we have shown that there must be gapped boundary theory of
a topological superconductor with $\nu'=\dim G-1$. For example, if we consider $G=E_8$ (which satisfies our condition $h^\vee \in 2\bZ$), we must get a topological theory
which reproduces the anomaly for odd $\nu$. It would be very interesting to investigate this direction in more detail.

\addtocontents{toc}{\protect\setcounter{tocdepth}{1}}

\bibliographystyle{ytphys}
\baselineskip=.9\baselineskip
\bibliography{ref}

\end{document}